# Constant-pressure sound waves in non-Hermitian disordered media

Etienne Rivet[1], Andre Brandstötter[2], Konstantinos G. Makris[3], Hervé Lissek[1], Stefan Rotter[2] and Romain Fleury[4, *]

[1] *Signal Processing Laboratory 2, EPFL, 1015 Lausanne, Switzerland*

[2] *Institute for Theoretical Physics, Vienna University of Technology (TU Wien), Vienna 1040, Austria, EU*

[3] *Department of Physics, University of Crete, 71003 Heraklion, Greece, EU*

[4] *Laboratory of Wave Engineering, EPFL, 1015 Lausanne, Switzerland*

*To whom correspondence should be addressed. Email: romain.fleury@epfl.ch*

**When waves impinge on a disordered material they are back-scattered and form a highly complex interference pattern[1-3]. Suppressing any such distortions of a wave's free propagation is a challenging task with many applications in a number of different disciplines[4-11]. In a recent theoretical proposal, it was pointed out that both perfect transmission through disorder as well as a complete suppression of any variation in a wave's intensity can be achieved by adding a continuous gain-loss distribution to the disorder[12-14]. Here we show that this abstract concept can be implemented in a realistic acoustic system. Our prototype consists of an acoustic waveguide containing several inclusions that scatter the incoming wave in a passive configuration and provide the gain or loss when being actively controlled. Our measurements on this non-Hermitian acoustic metamaterial demonstrate unambiguously the creation of a reflectionless scattering wave state that features a unique form of discrete constant-amplitude pressure waves. In addition to demonstrating that gain-loss additions can turn localised systems into transparent ones, we expect our proof-of-principle demonstration to trigger interesting new developments not only in sound engineering, but also in other related fields such as in non-Hermitian photonics.**

The disordered structure of a material renders many of the objects in our everyday lives opaque (we cannot see through them) or impenetrable for sound (that is our hearing is impaired when we plug them in our ears). In contrast to crystalline structures that have well-defined transmission bands and corresponding gaps between them, a disordered medium leads to multiple scattering that typically prevents wave transmission in very broadband spectral intervals. This aspect constitutes, in fact, a central challenge for many active research areas such as for biomedical imaging[4], metamaterial design[5,11], wireless data transfer[6], and wave control[8-10]. The strategies available to cope with the adverse effects of disorder scattering in these different domains typically involve either shaping the wavefront that impinges on the disorder[4,6,8-10], redesigning the internal structure of the disordered medium[7,11], exploiting topological protection[15], or cloaking the medium altogether[5]. A novel and potentially very useful approach to achieve novel functionalities of a system is to imprint on it a suitable distribution of gain and loss in the context of non-Hermitian wave physics[16-38], a research area initially driven by theoretical and experimental studies of non-Hermitian photonic systems[20-23] that respect Parity-Time symmetry[16-17]. Theoretical calculations[12-14] show that, in this way, so-called "constant-intensity (CI) waves" can be generated that perfectly penetrate even through disordered media without any backscattering and intensity variations throughout the entire transmission process. In other words, such CI waves have the very unconventional feature that they can propagate even through strongly disordered structures like a plane wave through free space. The only signature they carry from the inhomogeneous medium that they penetrate is in the phase they accumulate, but not in their intensity.

As we demonstrate here, these exotic wave states do not just exist for light, as originally predicted[13], but also for airborne sound waves. This property opens up the exciting possibility to realise CI waves in an acoustic wave system, for which, contrary to our everyday experience, sound propagates through a non-uniform medium without any variations in its pressure amplitude. To illustrate this conceptually, we firstly consider a finite one-dimensional scattering geometry of length $2L$ embedded in a uniform background acoustic medium. In the scattering region between $-L$ and $L$, the bulk modulus $\kappa(x)$ and the complex mass density are position-dependent such as

$$\rho(x) = \rho_R(x) + i\rho_I(x). \qquad (1)$$

The acoustic pressure $p(x)$ follows the acoustic Helmholtz equation

$$[\Delta + \omega^2 \rho(x) \kappa^{-1}(x)] p(x) = 0, \qquad (2)$$

where $\Delta = d^2/dx^2$ is the Laplacian in 1D and $\omega = 2\pi f$ is the angular frequency with $f$ being the frequency. We now make an ansatz for an acoustic wave $p(x)\exp(i\omega t)$ traveling only in positive $x$-direction with constant pressure and a position-dependent phase between $-L$ and $L$

$$p(x) = A\, e^{ik \int_{-L}^{x} W(x')\,dx'}, \qquad (3)$$

where $A$ is the constant (spatially independent) amplitude of the pressure, $k$ is the wavenumber, and $W(x)$ is a real auxiliary function, which can be chosen arbitrarily. Inserting Eq. (3) into Eq. (2) provides us with a design principle for the real and imaginary parts of the mass density,

$$\rho_R(x) = \frac{\kappa(x) k^2 W^2(x)}{\omega^2} \qquad (4)$$

$$\rho_I(x) = -\frac{\kappa(x) k W'(x)}{\omega^2} \qquad (5)$$

with $W' = dW(x)/dx$. Note that the wavenumber $k$ appearing in $\rho_R(x)$ and $\rho_I(x)$ in Eqs. (4,5) is the same as in the ansatz in Eq. (3), meaning that only waves with this specific wavenumber impinging from the left-hand side on this structure lead to a constant pressure wave also inside the structure as illustrated in Fig. 1. This approach can easily be generalized so as to also consider complex distributions of $\kappa(x)$.

For the trivial case that both the auxiliary function $W(x) = 1$ and the bulk modulus $\kappa(x) = \kappa$ are independent of position, we get $\rho_I(x) = 0$ and $\rho_R(x) = \kappa k^2/\omega^2$. Using $\omega = kv$, we get $v = \sqrt{\kappa/\rho}$ which is the well-known expression for the speed of sound in a uniform material. In this sense, our approach generalises the trivial plane wave solution $p(x) = A\exp(ikx)$ to the non-trivial case of an inhomogeneous medium, where acoustic gain and loss $[\rho_I(x) \neq 0]$ are required to maintain a spatially constant pressure wave.

A major obstacle to the practical implementation of this elegant theoretical concept is the challenge of building continuous gain-loss distributions in realistic systems. To bring this abstract idea to life, we extend the notion of constant-intensity scattering states to systems involving discontinuous and discrete distributions of gain and loss. Consider for this purpose a one-dimensional acoustic metamaterial composed of an air-filled tube, loaded with a set of discrete acoustic inclusions modelled by acoustic impedances $Z_j$. At low frequencies, where only a single mode can be excited, the system can be described by a transmission-line (TL) model, as illustrated in Fig. 2a. According to acoustic TL theory, the line voltage represents the complex acoustic pressure $p$, and the line current represents the volume flow $q$[39]. The black boxes represent the acoustic inclusions, introducing some discontinuities of the acoustic pressure along the line, while preserving the volume flow. The grey subsystems correspond to arbitrary transfer matrices $M_j = [A_j, B_j; C_j, D_j]$ that connect these impedances. It is important to note that in this description, non-Hermiticity is represented by the non-zero *real* part of the impedance $Z_j$ (describing positive or negative value of resistance), whereas the *imaginary* part corresponds to the Hermitian component (describing capacitive or inductive behaviour). Indeed, the imaginary and real parts of these impedances correspond to local modifications of the real and imaginary parts of the acoustic density, respectively[35]. We assume that the finite system with altogether $n$ inclusions is connected to two semi-infinite half-spaces described by characteristic impedances $Z_L$ (left) and $Z_R$ (right).

As we demonstrate here explicitly, such an acoustic system can support a powerful, discrete version of the constant-pressure sound wave. As in the continuous case, we start by considering an incoming wave with unit amplitude from the left and assuming that the acoustic pressures $p_j$ only differ in their phases, but not in their amplitudes, enforcing $p_{j+1} = e^{ik\varphi_j}p_j$ at the sites $j \in [\![1,n]\!]$. In Methods, we show that this assumption forces the volume flows $q_j$ to take the values

$$q_j = \frac{1}{\prod_{l=1}^{j-1} D_l} q_1 - \sum_{l=1}^{j-1} \frac{C_l}{\prod_{m=l}^{j-1} D_m} p_{l+1}. \tag{6}$$

The required acoustic impedance at the site $j$ is then found to be

$$Z_j = \frac{p_j - A_j p_{j+1} - B_j q_{j+1}}{C_j p_{j+1} + D_j q_{j+1}}. \tag{7}$$

Equation (7) directly implies that in order to obtain a constant-pressure sound wave at every site $j \in [\![1, n+1]\!]$, the real parts of the corresponding acoustic impedances $Z_j$ need to be non-zero in general. Therefore, the non-Hermiticity of a system is, indeed, a basic requirement for the realization of non-trivial constant-amplitude waves.

To illustrate the unique properties of these wave states, we consider the example of a disordered acoustic system that is initially Hermitian and terminated by $Z_L = Z_R =$ $Z_0$ (altogether $n = 8$ inclusions are incorporated). We also assume that they are connected by identical air-filled tube portions whose transfer matrices are defined by $A_j = D_j = \cos kd$, $B_j = iZ_0 \sin kd$, $C = iZ_0^{-1} \sin kd$, where $Z_0$ is the characteristic acoustic impedance in the tube. In this case, the corresponding acoustic impedances $Z_j$ are first assumed to be purely imaginary, corresponding to lossless elements that are either pure masses or springs. Their impedance values are randomly chosen from a uniform distribution. Figure 2b illustrates the impedance at each inclusion by a point on the complex plane, whose colour is associated with the considered site. Upon unitary excitation from the left, the acoustic pressure illustrated in Fig. 2c is strongly non-uniform at the sites $[\![1, n+1]\!]$, and its amplitude is very small at the output (yellow point). This case illustrates how disorder usually prevents efficient wave transmission.

We compare this situation with the case of a non-Hermitian system, whose acoustic impedances $Z_j$ have the same imaginary parts as in the previous case, but now their real parts are non-zero, following Eq. (7). As shown in Fig. 2d, and comparing with Fig. 2b, each point has now moved away from the imaginary axis along the horizontal direction, which corresponds to the punctual addition of loss or gain to the system. As can be seen in Fig. 2e, these non-Hermitian modifications have a significant impact on the distribution of the acoustic pressures within the system. Remarkably, all the acoustic pressures at the sites $[\![1, n+1]\!]$ now stick to the unit circle, such that they have the same amplitude as the incident wave, and different phases that exactly match the values for the phases $\varphi_j$ chosen by design. As a result, the pressure wave gets perfectly transmitted.

In order to better grasp the microscopic field dynamics related to this extraordinary phenomenon, including the behaviour of the pressure in between the non-Hermitian inclusions, we simulated a 3D air-filled cylindrical pipe loaded with eight transverse membranes ($n = 8$), whose corresponding acoustic impedances are set by applying appropriate internal impedance boundary conditions. Figure 3a shows the real part of the acoustic pressure (at a given moment of time; black line) obtained for excitation from the left for the case of the purely Hermitian system with a large degree of disorder (values given in Fig. 2b). All the sites illustrated in colour are lossless and correspond to purely reactive but mismatched acoustic impedances (with $Z_{n+1} = 0$, consistent with our convention in Fig. 2a). The grey area represents the values spanned by the pressure at each point over a full period of the harmonic field. As a result of the large disorder present in the system, the acoustic wave is mostly reflected and the absolute value of the pressure decreases rapidly along the system. In a next step this situation is compared to the corresponding non-Hermitian system designed to support a constant-pressure wave. In Fig. 3b, the eight inclusions are tuned to provide the right amount of gain and loss as calculated by our theory (consistent with Fig. 2d). Remarkably, the acoustic pressure now propagates through the system and the wave is perfectly transmitted at site $n + 1$. In addition, we notice that the grey envelope meets the unity dashed red line right in front of every membrane inclusion corresponding to the location of every pressure site $p_j$. In contrast to continuous constant-amplitude waves, the discrete design adopted here forces the pressure amplitude to be unitary at each check point $p_j$, but not in between. Notice, however, that this is not a severe limitation of the concept, since the distance between neighbouring sites $p_j$ can be chosen arbitrarily small and, when being subwavelength, prevents large amplitude fluctuations in between the sites. In addition, since the pressure amplitude at the last site $p_{n+1}$ is fully controlled, despite the fact that no inclusion is present at that location, perfect transmission can be guaranteed by design. We conclude that the present discretisation of the constant amplitude property does not introduce any fundamental limitation or change in functionality when compared to the continuous case, while it brings a significant simplification on the required gain-loss distribution. Our findings therefore suggest that a proper discrete distribution of gain and loss offers sufficient control over the scattering state in a strongly disordered system, thanks to local injection and absorption of energy within the material.

So far, we have only demonstrated a special case where the membrane inclusions were at the same time the source of Hermitian disorder and of the required gain and loss. However, our theory is more general and can accommodate any type of disorder placed at any location, including away from the pressure sites. Indeed, the transfer matrices $M_j$, so far considered constant and identical at all sites, can in general be different and describe the presence of scattering defects of any type. To prove this, we consider the full-wave scattering scenario of Figure 4a. An acoustic pipe is filled with defects of various kinds, including pipe section discontinuities, hard-wall obstacles, porous walls, labyrinthine detours, and even a strongly resonant side Helmholtz resonator. These defects, by themselves, completely prevent efficient wave transmission through the system, as shown in Fig. 4a. Yet, knowing their transfer matrices $M_j$, one can use our theory [Eq. (7)] to determine proper values of gain and loss to add between these defects, in order to turn this opaque system into a transparent one (Fig. 4b). From the figure it is clear that at each pressure site $p_1$ to $p_7$ the amplitude of the pressure has been restored to unity, providing an interesting way of steering sound seamlessly through an arbitrary combination of obstacles by engineering a special scattering state with discrete constant amplitude.

Based on our discrete theory, we built a non-Hermitian metamaterial prototype as shown in Fig. 5. This 2.75 m-long acoustic metamaterial consists of eight identical loudspeakers placed between identical cylindrical acoustic waveguide sections of length $d$. In the low-frequency approximation, the electrodynamic loudspeaker can be modelled as a one-degree-of-freedom oscillator (that is a mass-spring-damper system) mechanically driven by a voice coil within a magnetic field. When an appropriate electrical circuit is connected to the transducer terminals, it is possible to control the dynamic mechanical behaviour of the diaphragm[40]. Here, each loudspeaker is connected to a control unit designed to assign a target acoustic impedance [given in Eq. (7)] to the diaphragm at the design frequency. This control unit regulates the diaphragm velocity of the current-driven loudspeaker at site $j$ by measuring the front and rear sides acoustic pressures $p_{fj}$ and $p_{rj}$, and returning the proper electrical current according to the target complex gains (see Methods). A sound source is located at the left side of the system and is connected to a tube of cross-section area $S$. A tube of the same cross-section area is connected at the right side of the system, corresponding to a situation in which $Z_L = Z_R = Z_0$. Lastly, the rightmost end of the system is terminated by a matched load that guarantees no reflection at the design frequency (121 Hz). Figures 5a and 5b show a photograph and the schematic of the complete set-up, respectively.

Figure 6 reports our measurements (coloured data), comparing it with theory (black data). First, we considered a reference Hermitian system in which the imaginary parts of acoustic impedances where chosen in the interval $[-1.8\,Z_0, 0]$ (negative values of reactance are more easily obtained from the loudspeakers at the frequency of operation, due to the bulk of their driver). Statistical considerations based on ensemble averages show that such a system operates in the regime of Anderson localisation, with a localisation length just below one meter, making the total prototype about three times the localisation length (see Methods). The realisation of disorder considered in the experiment is shown in Fig. 6b. The acoustic pressures that were measured directly in front of the inclusions with the help of microphones are displayed in Fig. 6c. Clearly, the localised scattering state has a spatially very inhomogeneous amplitude, and a strongly reduced transmission to the other side (measured at 33%). Conversely, when we turn on the control by adding gain and loss in amounts prescribed by theory (Fig. 6d), the measured pressure amplitudes become identical at each site (Fig. 6e), demonstrating the unique ability of non-Hermitian systems to fight disorder. In this situation, we measured a unitary transmittance (with percent-level accuracy), and near-perfect agreement with theory. The very small discrepancies found did not affect the constant-amplitude property, which was observed to be quite robust to small variations of the impedances. Altogether, our measurements unambiguously demonstrate that a tailor-made distribution of gain and loss opens up a transparency window in a disordered and Anderson-localised Hermitian system in which discrete pressure waves can propagate with constant-amplitude.

In conclusion, we have put forward and experimentally demonstrated the concept of discrete constant-amplitude sound propagation in non-Hermitian systems. Such waves are the generalisation of plane waves to disordered non-Hermitian systems. Our theory of discrete, non-Hermitian acoustic systems with waves of constant pressure is exact, very general, and does not rely on some form of homogenisation procedure or unpractical continuous gain-loss distributions. Our experimental results demonstrate that non-Hermiticity can be used to counteract the detrimental effects of disorder even in Anderson-localised systems. Specifically, we locally control the power generation and absorption at the microscopic scale and thereby engineer the acoustic pressure to have desired values at discrete check points. Our concept offers interesting technological perspectives by enriching the toolkit of acousticians with a solution for acoustic transmission through multiple-scattering media that are conventionally opaque. In addition, our experiment provides a beautiful tabletop platform offering fascinating insights into the physics of non-Hermitian disordered systems.

**Figures**

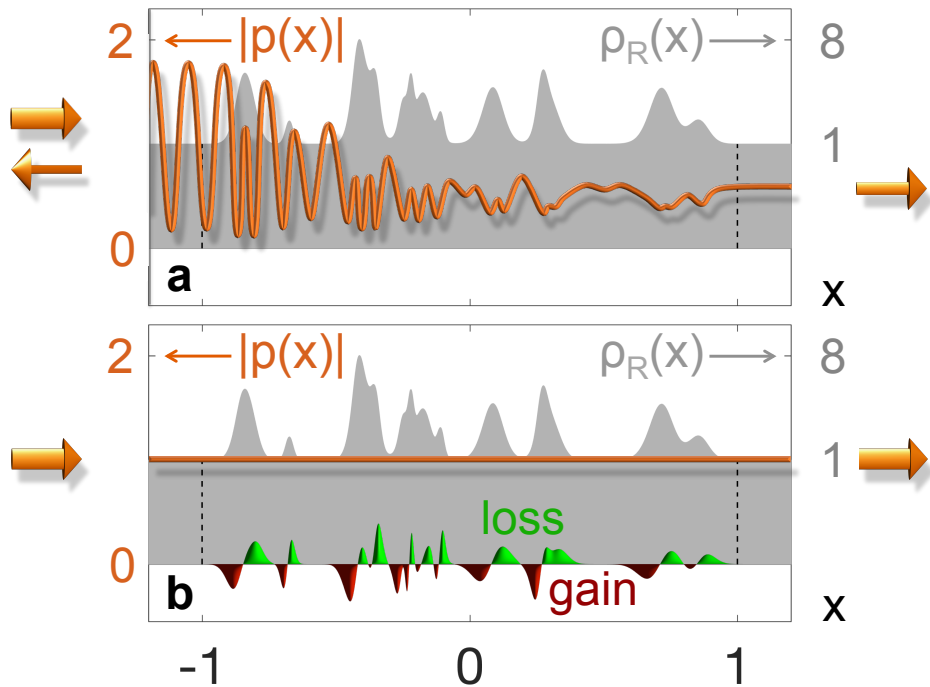

**Figure 1: Concept of continuous constant-pressure waves. a)** Absolute value of the acoustic pressure $p(x)$ (orange line) in a disordered Hermitian system when a wave is incident from the left-hand side with normalized wavenumber $k = 2\pi/0.27$. Due to variations of the mass density $\rho_R(x)$ (grey), the acoustic pressure wave is partially reflected and shows strong variations inside the medium. **b)** Adding a tailored gain-loss distribution (red/green) according to Eqs. (4,5) to the system shown in **a),** the acoustic pressure field gets perfectly transmitted and shows no pressure variations inside the entire scattering region.

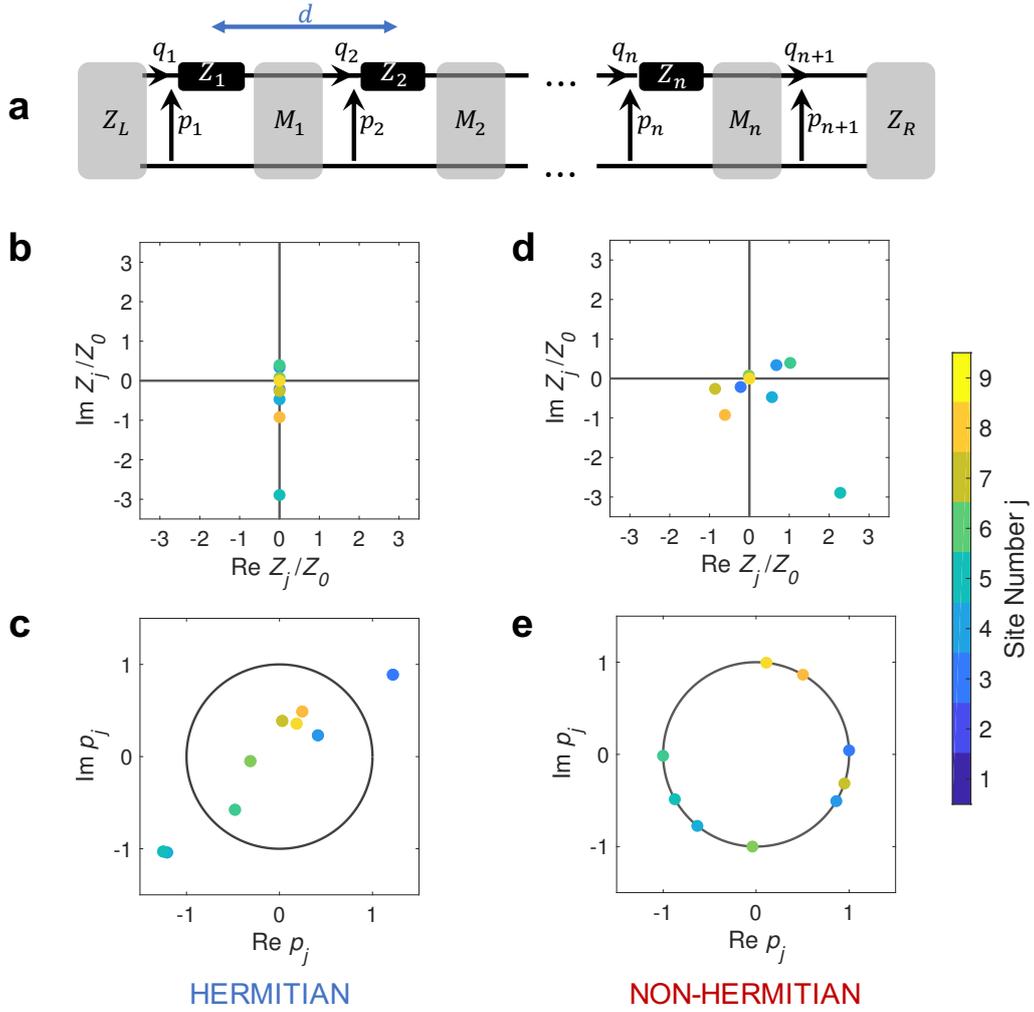

**Figure 2: Discrete constant-pressure acoustic waves. a)** We consider an acoustic transmission-line of characteristic impedances $Z_L$ at the left end and $Z_R$ at the right end, loaded with $n$ series impedances $Z_j$. Each imaginary part of $Z_j$ is used to introduce Hermitian disorder to the system, whereas the real part implements an additional discrete distribution of loss or gain. The inclusions are connected by transfer matrices $M_j$, which may correspond to additional disorder in general, but are assumed to be identical ($M_j = M_0$) in panels b-e, with $M_0$ representing a simple Hermitian portion of transmission line with impedance $Z_0$ and length $d$. We first consider the particular case of a Hermitian system ($n = 8$) with random but purely imaginary impedances as shown in **b)**. **c)** The acoustic pressures $p_j$ obtained from our analytical model for the case in **b)** have non-uniform amplitudes. **d)** By adding loss and gain, we create a non-Hermitian system whose non-Hermitian impedances have the same imaginary parts as those in the case of the Hermitian system **b)**, but their real parts are now different from zero. **e)** Tailoring these non-Hermitian additions results in an acoustic scattering state with constant-amplitude pressure at each site $p_j$.

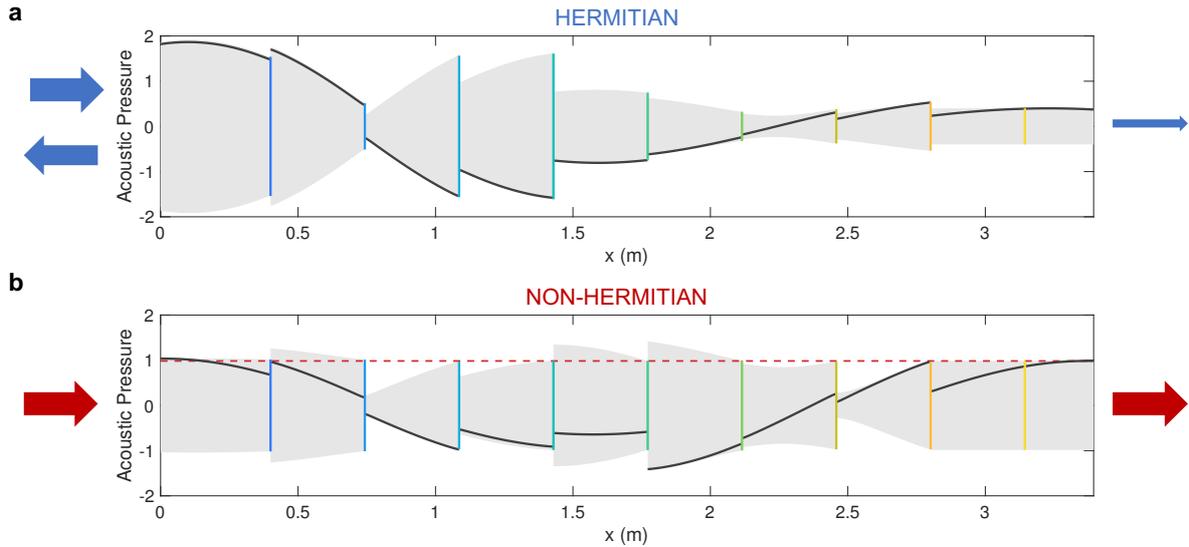

**Figure 3: Microscopic field dynamics in a non-Hermitian metamaterial supporting discrete constant-pressure acoustic waves.** We performed full-wave 3D finite-element simulations to investigate the microscopic behaviour of the pressure field in the Hermitian case of Fig. 2b-c (panel **a**) and the corresponding non-Hermitian case of Fig. 2d-e (panel **b**). The black line is a snapshot in time of the acoustic pressure field, whereas the shadowed area shows the range of its oscillations as time evolves. The vertical coloured lines represent the locations of the impedances $Z_j$. **a)** The disordered Hermitian system transmits poorly and the amplitude inside fluctuates strongly. **b)** Adding gain or loss to each inclusion counteracts the effect of disorder and makes the system transparent. The discrete constant amplitude property implies that the amplitude of the pressure reaches unity at each check point $p_j$, located directly in front of each inclusion (consistent with Fig. 2a). In panel **b)**, the shadowed grey area indeed touches the dashed red unity line right before each impedance.

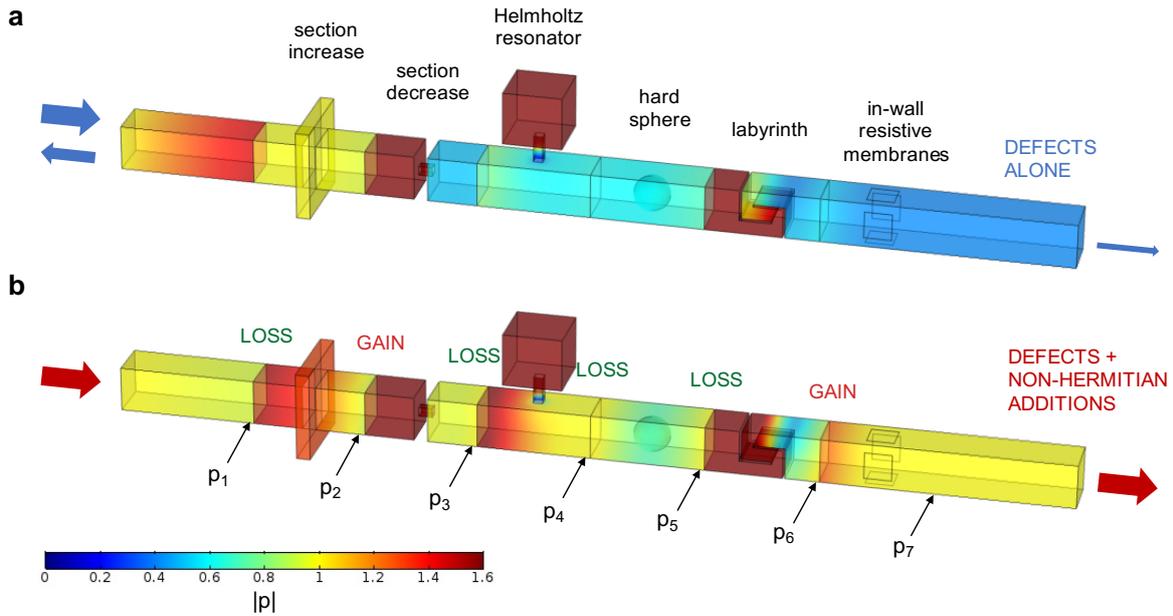

**Figure 4: Adding gain and loss to fight scattering defects of any type. a)** We consider a pipe filled with various strongly scattering obstacles that prevent efficient transmission of sound (colour map shows pressure amplitude). We have introduced pipe section discontinuities, a resonant side Helmholtz resonator, a hard wall obstacle, a labyrinthine path and porous walls. When only the defects are considered, the pressure amplitude is fluctuating, and reaches almost zero on the transmission side. **b)** After adding the discrete distribution of gain and loss as prescribed by our theory, we obtain perfect sound transmission. The pressure amplitudes are constant at discrete locations right in front of each added non-Hermitian inclusion (controlled sites $p_1$ to $p_7$).

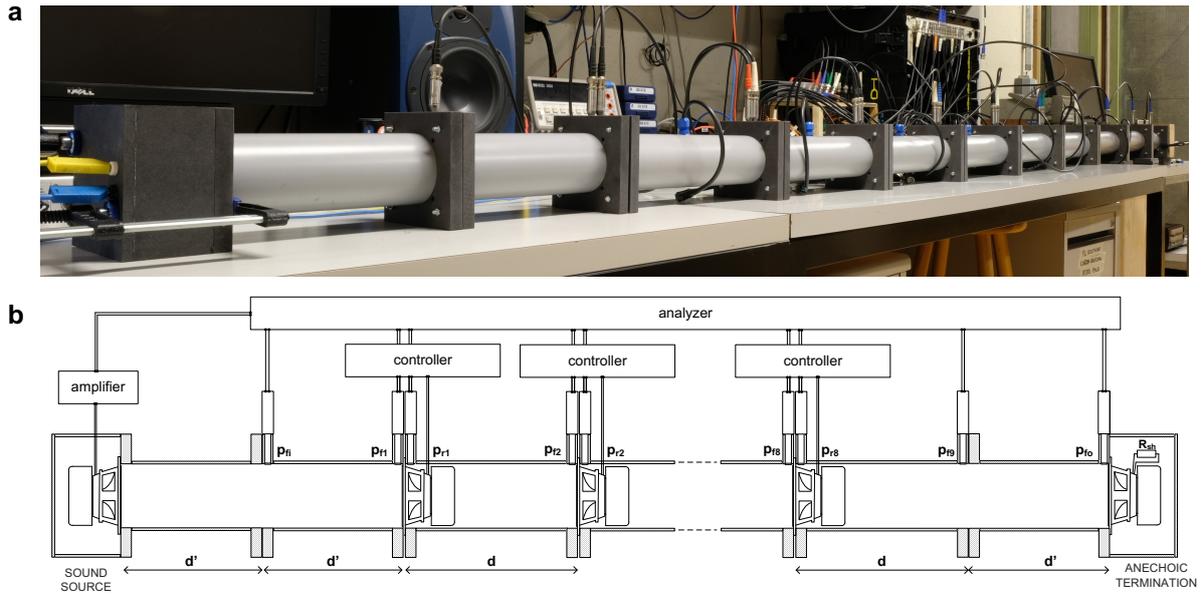

**Figure 5: Non-Hermitian metamaterial prototype supporting discrete constant-amplitude pressure waves. a)** We built a one-dimensional acoustic metamaterial consisting of a 3.5 m-long waveguide loaded with eight non-Hermitian acoustic inclusions ($8d = 2.75$ m). The inclusions are electrodynamic loudspeakers, whose acoustic impedance can be conveniently tailored electrically through a harmonic impedance control method. **b)** Schematic of the experimental set-up showing the inclusions with the control, the sound source at the left end, and the anechoic termination at the right end.

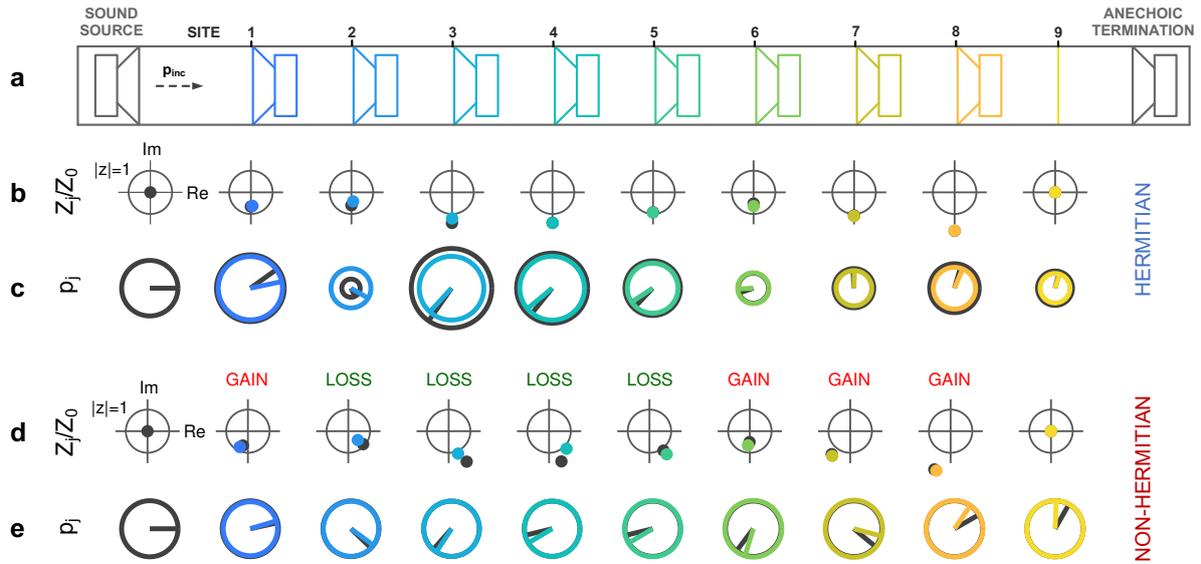

**Figure 6: Experimental proof of discrete constant-amplitude pressure waves.** We performed two experiments: a reference one with a purely Hermitian disordered system (panels b-c), and a second one on the same system with added gain and loss according to the theory (panels d-e). Measured values are shown with colour lines and predicted values with black lines. **a)** Schematic of the setup, serving as a column reference for the other panels. **b)** Measured normalized acoustic impedances in the Hermitian case represented in the complex plane. All impedances are imaginary, unambiguously demonstrating the Hermitian origin of the disorder. **c)** Pressures measured at each site in the Hermitian case. Pressures are represented as phasors, with the radius of the circle being proportional to their amplitude, and the line segment showing their phase with respect to that of the incident pressure. **d)** Measured impedances in the non-Hermitian case, highlighting the addition of gain or loss to each inclusion. **e)** Pressures measured at each site in the non-Hermitian case, demonstrating the discrete constant amplitude property in excellent agreement with theoretical predictions.

## Methods

**Discrete constant-pressure waves.** The situation represented in Fig. 2a can be analysed through the transfer matrix formalism[41]. The pressure $p_j$ and volume flow $q_j$ at each inclusion $j \in [\![1, n]\!]$ are related by:

$$\begin{pmatrix} p_j \\ q_j \end{pmatrix} = \begin{pmatrix} 1 & Z_j \\ 0 & 1 \end{pmatrix} \begin{pmatrix} A_j & B_j \\ C_j & D_j \end{pmatrix} \begin{pmatrix} p_{j+1} \\ q_{j+1} \end{pmatrix}, \qquad (8)$$

where $M_j = [A_j, B_j; C_j, D_j]$ represents the transfer matrices connecting the impedances $Z_j$ and $Z_{j+1}$. In addition, assuming an incident wave from the left with amplitude $p_{inc} = 1$ Pa, we use the boundary conditions

$$p_{n+1}/q_{n+1} = Z_R \qquad (9)$$

$$p_1/p_{inc} = 1 + R \qquad (10)$$

$$q_1/p_{inc} = (1 - R)/Z_L, \qquad (11)$$

where $R$ is the reflection coefficient. If we assume that the acoustic pressures have the same amplitude at all the sites, enforcing $p_{j+1} = e^{ik\varphi_j} p_j$, we have $p_{j+1} = \exp\left(ik \sum_{l=1}^{j} \varphi_l\right) p_1$. Then, all the pressures $p_j$ are known as a function of $p_1$, or equivalently, $R$ [see Eq. (10)]. This choice for the pressures $p_j$ also fixes the corresponding volume flows $q_j$ according to Eq. (8). Indeed, solving the recurrence equation $q_j = C_j p_{j+1} + D_j q_{j+1}$ directly yields Eq. (6). Then, with Eqs. (6) and (11), we also know all the volume flows $q_j$ as a function of the reflection coefficient $R$. To determine the value of $R$, we need to enforce the boundary condition at the end of the line given in Eq. (9). We obtain

$$(1 + R) e^{ik \sum_{l=1}^{n} \varphi_l} = Z_R \left( \frac{1}{\prod_{l=1}^{n} D_l} q_1 - \sum_{l=1}^{n} \frac{C_l}{\prod_{m=l}^{n} D_m} p_{l+1} \right) \qquad (12)$$

which yields, $R = (Z_{in} - Z_L)/(Z_{in} + Z_L)$ where $Z_{in}$ is the input acoustic impedance (at site 1) expressed as

$$Z_{in} = Z_R \frac{1}{\prod_{l=1}^{n} D_l} \frac{1}{e^{ik \sum_{l=1}^{n} \varphi_l} + Z_R \sum_{l=1}^{n} \frac{C_l}{\prod_{m=l}^{n} D_m} e^{ik \sum_{r=1}^{l} \varphi_r}}. \qquad (13)$$

Since all the acoustic quantities are completely determined, we can calculate the target acoustic impedances $Z_j$ from Eq. (8), directly getting Eq. (7). We see that if we set all the pressure phases $\varphi_j$, then both the acoustic impedances $Z_j$ and the reflection coefficient $R$ are fixed. Another possibility is to set $R = 0$ to obtain a system without any reflection, at the cost of relaxing two degrees of freedom in the choice of the phases. For instance, we can impose the sum of the phases to be constant (in order to be able to set the total phase of the transmission coefficient) and tune two values of phase to enforce the reflection coefficient $R = 0$. Since the impedance $Z_j$ at the inclusion $j$ in Eq. (7) does not depend on the pressure and volume flow of previous sites $[\![1, j-1]\!]$ on the TL, we used in our numerical simulations the first two phases $\varphi_1$ and $\varphi_2$ to set $R = 0$ without modifying the rest of the design. Using this theory, we can generate many matched systems supporting constant-amplitude pressure waves, even for arbitrarily large values of $n$. Note that even in the case $R \neq 0$, our theory leads to a distribution of gain and loss in the system with constant pressure at every site $j \in [\![1, n+1]\!]$.

**Harmonic acoustic impedance control.** In the low-frequency approximation, the electrodynamic loudspeaker can be modelled as a one-degree-of-freedom oscillator mechanically driven by a voice coil within a magnetic field. All forces acting on the transducer, especially those resulting from the sound pressures $p_f$ and $p_r$ at the front and rear faces of the diaphragm, are assumed small enough so that the governing equations remain linear. The mechanical part is assumed as a simple mass - spring - damper system in the low-frequency range, that is the mass $M_m$, the mechanical compliance (inverse of stiffness) $C_m$ accounting for the surround suspension and spider, and the mechanical resistance $R_m$, respectively. If we denote the effective piston area by $S_d$ and the force factor of the moving-coil transducer by $Bl$, the equation of motion of the loudspeaker diaphragm at inclusion $j$ is derived from Newton's second law, which can be written using the Fourier transform as

$$S_d P_{fj}(\omega) - S_d P_{rj}(\omega) = Z_{mj}(\omega) V_j(\omega) + Bl_j I_j(\omega), \tag{14}$$

where $Z_{mj}(\omega) = i\omega M_{mj} + R_{mj} + 1/(i\omega C_{mj})$ is the mechanical impedance of the loudspeaker, $V_j(\omega) = Q_j(\omega)/S_d$ is the diaphragm velocity, and $I_j(\omega)$ is the current flowing through the voice coil of the loudspeaker.

Our control strategy uses two microphones giving the signals $\widehat{P_{fj}}(\omega) = \sigma_{fj} P_{fj}(\omega)$ and $\widehat{P_{rj}}(\omega) = \sigma_{rj} P_{rj}(\omega)$, where $\sigma_{fj}$ and $\sigma_{rj}$ are the sensitivities of the microphones (in V.Pa⁻¹) at the front and rear sides of the corresponding loudspeaker diaphragm respectively, and return the current $I_j(\omega)$ with transfer functions $\Theta_j(\omega)$[40] such as

$$I_j(\omega) = \Theta_j(\omega, \sigma_{fj}) \widehat{P_{fj}}(\omega) - \Theta_j(\omega, \sigma_{rj}) \widehat{P_{rj}}(\omega). \tag{15}$$

Assuming the target acoustic impedance $Z_{tj}(\omega)$, which is defined in Eq. (7), is realized at the diaphragm, the transfer function $\Theta_j(\omega)$ from the microphone signal $\widehat{P_{fj}}(\omega)$ or $\widehat{P_{rj}}(\omega)$ to the electrical current $I_j(\omega)$ can be derived from Eqs. (14) and (15) as

$$\Theta_j(\omega,\sigma) = \frac{S_d}{\sigma Bl_j}\left(1 - \frac{Z_{mj}(\omega)}{S_d^2 Z_{tj}(\omega)}\right). \qquad (16)$$

It is worth noting that the signal filtering through the transfer function $\Theta_j$ should be processed before differentiating the signals in Eq. (15), because of possible loss of significance in the controller. Note also that the situation $\Theta_j(\omega,\sigma) = 0$ through the voltage-controlled current source corresponds to the case where the loudspeaker is in open circuit [see Eqs. (14) and (15)].

For the inclusions giving gain to the system, that is $\text{Re}(Z_{tj}) < 0$ at the frequency of interest $f_0$, the transfer function $\Theta_j(\omega)$ is in this case not causal and therefore, it cannot be implemented in a digital controller. We thus turn the broadband impedance control[38] into a harmonic control through the formalism of demodulation/modulation[40]. First, the microphone signals $\widehat{P_{fj}}(\omega)$ and $\widehat{P_{rj}}(\omega)$ are demodulated at the angular frequency of interest $\omega_0 = 2\pi f_0$ (translating both signals in the frequency domain from $\omega_0$ to 0), whose low-pass filter of cut-off angular frequency $\omega_c \ll \omega_0$ is equal to 2 for $|\omega| < \omega_c$ and 0 otherwise. These complex signals are then multiplied by the target complex gain $\Theta_j(\omega_0,\sigma_{fj})$ or $\Theta_j(\omega_0,\sigma_{rj})$. The signals are then modulated at the angular frequency $\omega_0$ (translating both signals in the frequency domain from 0 to $\omega_0$). Lastly, the output signal of spectral range $[\omega_0 - \omega_c, \omega_0 + \omega_c]$ delivered by the voltage-controlled current source corresponds to the difference of both modulated signals as given in Eq. (15). This way, the target acoustic impedance $Z_{tj}$ is perfectly assigned at the loudspeaker diaphragm at the frequency of interest.

**Experiment.** The experimental setup consisted of eight electrodynamic loudspeakers ($n = 8$), separated from each other by a distance $d = 34.3$ cm in a tube of cylindrical cross-section area $S = \pi r^2 = 40.7$ cm$^2$ (see Figure 5 and Supplementary Figure 1). At the left termination we placed a sound source (closed-box loudspeaker) that delivered a band-limited sweep sine of bandwidth [20 Hz – 400 Hz] with the help of a Brüel & Kjaer 2706 power amplifier. At the other termination we placed a matched load with the help of another closed-box loudspeaker. A mass was added to the diaphragm and a resistive shunt was connected to the electrical terminals[43] to decrease the resonance frequency and match the acoustic resistance at the diaphragm to $Z_0$ respectively, guaranteeing no reflection at the design frequency (121 Hz). We only used Monacor SPX-30M loudspeakers of effective piston area $S_d = \pi r_d^2 = 32$ cm² in the experiment. The first site was located at a distance $2d' = 50$ cm from the sound source and the last site at a distance $d'$ from the matched load. The target complex gain was defined for each controlled loudspeaker to get the exact values of acoustic resistance and reactance for each site given in Eq. (7) at the desired frequency. Eight pairs of ¼'' PCB 130D20/130F20 microphones were wall-mounted in front and behind every loudspeaker diaphragm. Each pair of microphones measured the acoustic pressures $p_{fj}$ and $p_{rj}$ at each site $j$. Three additional microphones were also used in the experiment. One microphone was located at around a distance $d'$ from the sound source, another one at the position of the ninth inclusion, and the last one in front of the matched load. The frequency responses between microphone signals at the different sites were processed through a Brüel & Kjaer Pulse 3160 multichannel analyser, so as to estimate the normalized acoustic impedance at every inclusion and the transmission coefficient. The control was implemented onto a real-time National Instrument CompactRIO-9068 platform supporting field-programmable gate array technology. The voltage signals from the eight pairs of microphones were digitally converted thanks to four analog modules NI 9215. The output signals were then delivered by two analog modules NI 9263 to voltage-controlled current sources that drove the loudspeakers.

**Estimation of the localisation length.** We have checked numerically that our disordered system is indeed operated in the localised regime, and that its size is about three times the localisation length. To verify this, we have computed the ensemble average of the logarithm of the transmission coefficient $<\ln T>$ over 1000 realisations of Hermitian disorder, for systems composed of an increasing number of sites (hence increasing the total length of the system). Importantly, the disorder strength was set to be the same as in the experiment (i.e. the normalized reactances were drawn randomly in the range $[-1.8, 0]$). We found that the quantity $<\ln T>$ varies linearly with the number of sites (varying from 4 to 20 unit cells, see supplementary Figure 2), which is the hallmark for 1D Anderson localisation. The inverse slope, which gives the Anderson localisation length, is extracted to be 2.7 inclusions (92.6 cm). Thus, our 2.75 m-long prototype is then comparable to three times the localisation length. Our experiment therefore clearly demonstrates unambiguously that the addition of gain and loss can turn a localised system into a transparent one.

**Measurement of acoustic impedance of every inclusion.** The bulkiness of each loudspeaker driver in the tube locally alters the pressure and volume flow behind the diaphragm. To take this acoustic behaviour into account in the design, we model the geometry of the loudspeaker using the transfer matrix formalism. As illustrated in Supplementary Figure 1 and according to Eq. (8), we choose $Z_j = Z_{mj}/S_d^2$ and

$$M_j = \begin{pmatrix} A_j & B_j \\ C_j & D_j \end{pmatrix} = M_p(Z_0, d_1) M_p(Z_2, d_2) M_p(Z_0, d - d_1 - d_2), \quad (17)$$

where

$$M_p(Z, x) = \begin{pmatrix} \cos(kx) & iZ\sin(kx) \\ i\sin(kx)/Z & \cos(kx) \end{pmatrix} \quad (18)$$

and $d_1 = 2.6$ cm, $d_2 = 2.8$ cm, $Z_2 = S/(S - S_2)Z_0$ with $S_2 = \pi r_2^2 = 37.4$ cm$^2$.

First, we consider a preliminary setup consisting of only one inclusion as illustrated in Supplementary Figure 1. Regardless of the terminations of this piece of waveguide, the acoustic impedance at the diaphragm may be estimated either from the acoustic pressures $p_A$, $p_B$, and $p_C$ (hereafter denoted as method 'ABC') such that

$$(Z_j)_{ABC} = iZ_0 \sin(kd) \frac{p_B - p_C}{p_A - \cos(kd) p_B} \quad (19)$$

or from the acoustic pressures $p_B$, $p_C$, and $p_D$ (method 'BCD') such that

$$(Z_j)_{BCD} = B_j \frac{p_B - p_C}{D_j p_C + (B_j C_j - A_j D_j) p_D}. \quad (20)$$

The acoustic impedance $Z_j$ can then be measured from Eqs. (19) and (20) with frequency responses $H_{BA} = \widehat{p_A}/\widehat{p_B}$, $H_{BC} = \widehat{p_C}/\widehat{p_B}$, and $H_{BD} = \widehat{p_D}/\widehat{p_B}$. When we performed these measurements for the preliminary setup, we found that the methods 'ABC' and 'BCD' gave consistent results. These methods were validated with a direct impedance measurement, from the microphone signals $\widehat{p_B}$ and $\widehat{p_C}$ and a Polytec OFV-5000 laser vibrometer. However, when all the loudspeakers are inserted between the tube sections in the acoustic metamaterial shown in Fig. 5, we can only use method 'BCD' (due to geometrical constraints for the inclusions [[2,8]]). With this method, the acoustic impedance of each inclusion was measured in-situ to estimate the model parameters of each loudspeaker through two measurements with a least squares fitting method. The first measurement was performed with the loudspeaker in open circuit to estimate the three parameters of the mechanical impedance $Z_{mj}$. The second measurement was performed with the loudspeaker in short circuit to estimate the force factor $Bl$. Then, every inclusion was measured again to verify that we were able to reach the target values of the non-Hermitian and Hermitian acoustic impedances. In Figs. 6b and 6d, we show at the frequency of interest the measured real and imaginary parts of each normalized acoustic impedance $Z_j/Z_0$ (colour dots) in the complex plane representation, in the Hermitian and non-Hermitian cases, respectively. These values are compared with the theoretical values (grey dots) that fulfil Eq. (7). The measurements confirm that the normalized acoustic impedances $Z_j/Z_0$ are very close to the theoretical ones.

## Data availability

The data that support the plots within this paper and other findings of this study are available from the corresponding author upon reasonable request.

## Author contributions

A.B., K.G.M., and S.R. developed the concept and theory of continuous constant amplitude waves. E.R. and R.F. developed the discrete theory of constant amplitude waves and performed the numerical simulations. E.R. formulated the acoustic impedance control theory, developed the control technology used in the experiment, and performed the experiment. H.L. supervised the experimental work. S.R. and R.F. initiated and supervised the project. All authors discussed the results and contributed to writing the manuscript.

## Acknowledgments

The authors would like to thank Prof. Mario Paolone and the Distributed Electrical Systems Laboratory at Ecole polytechnique fédérale de Lausanne for lending us the National Instrument CompactRIO-9068 platform for the experiment.